\documentclass[twoside,twocolumn]{article}

\usepackage{fitee}

\usepackage[colorlinks, breaklinks = true]{hyperref}		

\usepackage{amsfonts}
\usepackage{cite}
\usepackage{float}
\usepackage{graphicx}
\usepackage{amsmath}
\usepackage{times}
\usepackage{graphicx}
\usepackage{latexsym}
\usepackage{bm}
\usepackage{amssymb}
\usepackage{stfloats}
\usepackage{cases}
\usepackage{array}
\usepackage{setspace}
\usepackage{fancyhdr}
\usepackage{color}
\usepackage{multirow}
\usepackage{diagbox}
\usepackage{epstopdf}
\usepackage{varwidth}
\usepackage{pifont}
\usepackage{mathrsfs}
\usepackage{booktabs}
\usepackage{verbatim}
\usepackage{subfloat}
\usepackage{tabularx}

\addto{\captionsenglish}{%
  
}

\begin{document}

\title{Deep Unfolding-Based Channel Estimation for Wideband \\ Terahertz Near-Field Massive MIMO Systems}

\author[1]{Jia-bao GAO}%
\author[$\ddagger$1]{Xiao-ming CHEN}
\author[2]{Geoffrey Ye LI}
\affil[1]{College of Information Science and Electronic Engineering, Zhejiang University, Hangzhou 310027, China}
\affil[2]{Department of Electrical and Electronic Engineering, Imperial College London, London SW7 2BU, U.K.}

\shortauthor{Gao et al.}	

\authmark{}

\corremailA{gao\_jiabao@zju.edu.cn}
\corremailB{chen\_xiaoming@zju.edu.cn}
\corremailC{Geoffrey.Li@imperial.ac.uk}

\dateinfo{Received mmm.\ dd, xxxx;\	Revision accepted mmm.\ dd, xxxx;\ Crosschecked mmm.\ dd, xxxx;\ Published online mmm.\ dd, xxxx}

\abstract{The combination of Terahertz (THz) and massive multiple-input multiple-output (MIMO) is promising to meet the increasing data rate demand of future wireless communication systems thanks to the huge bandwidth and spatial degrees of freedom. However, unique channel features such as the near-field beam split effect make channel estimation particularly challenging in THz massive MIMO systems. On one hand, adopting the conventional angular domain transformation dictionary designed for low-frequency far-field channels will result in degraded channel sparsity and destroyed sparsity structure in the transformed domain. On the other hand, most existing compressive sensing-based channel estimation algorithms cannot achieve high performance and low complexity simultaneously. To alleviate these issues, in this paper, we first adopt frequency-dependent near-field dictionaries to maintain good channel sparsity and sparsity structure in the transformed domain under the near-field beam split effect. Then, a deep unfolding-based wideband THz massive MIMO channel estimation algorithm is proposed. In each iteration of the approximate message passing-sparse Bayesian learning algorithm, the optimal update rule is learned by a deep neural network (DNN), whose architecture is customized to effectively exploit the inherent channel patterns. Furthermore, a mixed training method based on novel designs of the DNN architecture and the loss function is developed to effectively train data from different system configurations. Simulation results validate the superiority of the proposed algorithm in terms of performance, complexity, and robustness.}


\doi{xx.xxxx/FITEE.xxxxxxx}	
\code{A}
\clc{TP}


\publishyear{xxxx}
\vol{1}
\issue{1}
\pagestart{1}
\pageend{1}

\support{The work was supported by the China National Key R\&D Program under Grant 2020YFB1805704.}

\orcid{Xiao-ming CHEN, http://orcid.org/0000-0002-1818-2135}	
\articleType{Science Letters}

\maketitle

\section{Introduction}
Terahertz (THz) massive multiple-input multiple-output (MIMO) is recognized as a promising technology in future wireless communication systems since the huge bandwidth and spatial degrees of freedom can support various emerging applications requiring high data rates \citep{THz_MIMO,mmWave_MIMO}. Nevertheless, this kind of appealing double gain heavily relies on the accuracy of available channel parameters, whose estimation is particularly challenging in THz massive MIMO systems due to several reasons.

Compressive sensing (CS) algorithms are able to recover high-dimensional channels from low-dimensional received pilots with reduced overhead. However, the performance of CS-based channel estimators will be severely degraded if the special features of THz massive MIMO channels are not properly handled. On one hand, the receiver will easily fall in the near field of the electromagnetic wave sent by the transmitter, as illustrated in Fig. \ref{system}. The boundary to divide the near- and the far-fields, defined as the Rayleigh distance, is proportional to the array aperture and inversely proportional to the wavelength \citep{near_field_dict}. Within the range of Rayleigh distance, spherical wavefront must be considered and simply using the approximately planar wavefront model will reduce the channel sparsity. In THz massive MIMO systems, the combination of short wavelength and large array aperture usually leads to Rayleigh distances of up to hundreds of meters, thus cannot be ignored in most cellular systems. On the other hand, with both large bandwidth and array aperture, the beam split effect appears such that the equivalent angles corresponding to the same physical channel path are different at different subcarriers \citep{near_field_beam_split}. As a result, the widely exploited common sparsity structure among subchannels no longer holds. Last but not least, the complexity of most existing high-performance CS algorithms becomes unbearable with massive antennas.

To deal with the near-field effect, the authors of \citep{near_field_dict} have derived the near-field array response vector geometrically, which is dependent on not only the angle but also the distance. Then, the near-field dictionary has been constructed by multiple near-field array response vectors with different angle and distance grids. The classic multiple-measurement-vectors (MMV) CS algorithm, simultaneous orthogonal matching pursuit (SOMP), has been adopted for channel estimation exploiting the common sparsity structure. In \citep{hybrid_far_near}, the hybrid-field scenario has been considered, where the far-field paths and the near-field paths are estimated sequentially. For the mixed line-of-sight (LoS) and non-line-of-sight (NLoS) scenario, the LoS path component has been first estimated in \citep{mixed_los_nlos} in an off-grid manner and then the NLoS path components have been processed in an on-grid manner. To promote the structured sparsity under the spatial non-stationary effect caused by the spherical wavefront and visibility region issue, the prior of the orthogonal approximate message passing (OAMP) algorithm is tailored in \citep{non_stationary}. To compensate for the beam split effect, dictionary is designed to be frequency-dependent in \citep{NB_OMP} so that the common sparsity structure can be maintained while the bilinear pattern detection method in \citep{bpd} collects powers of the strongest locations in the angle-distance domain from all frequencies.

Deep learning (DL) has achieved great success in many wireless communication problems in the past few years \citep{dl_survey}. Recently, DL has also been applied to near-field THz massive MIMO channel estimation. In \citep{NB_OMP}, a black-box deep neural network (DNN) has been used to predict the wideband THz near-field channels based on the received pilot signals while only the key parameters of channel paths including angles, distances, and gains are predicted in \citep{dcnn_thz_ce}. In \citep{refine_omp}, the coarse estimation of OMP is refined by a denoising autoencoder to further improve estimation performance. To reduce the complexity, in \citep{LISTA_near_field}, a smaller dictionary is learned together with parameters inserted into the iterative shrinkage and thresholding algorithm. To realize adaptive complexity and guarantee linear convergence, the efficient channel estimator in \citep{fixed_point_nn_near_field} is developed based on the fixed point DNN.

However, the above works either consider less practical narrowband scenarios or only have limited performance. Besides, most existing DL-based methods are trained separately under different system configurations, which increases the training and storage overhead and decreases the robustness. Therefore, estimators with high performance, low complexity, and strong robustness for practical THz massive MIMO channels still need to be investigated. In \citep{sbl_unfolding,uamp_sbl_unfolding}, the superiority of deeply unfolding the advanced (AMP-)sparse Bayesian learning (SBL) algorithm in wideband mmWave massive MIMO channel estimation has been validated. To extend this method to the THz band and enhance its robustness, we modify it in terms of dictionary design, network architecture, and training scheme in this paper. Our main contributions are summarized as follows:
\begin{itemize}
     \item We use frequency-dependent near-field dictionaries to compensate for the near-field beam split effect of wideband THz massive MIMO channels, thus improving the performance of the proposed DL-based channel estimator.
     \item We propose a deep unfolding-based channel estimation algorithm, where AMP is used to lower the complexity of the SBL algorithm and the DL-based parameter update procedure in each iteration improves both the convergence speed and the converged performance.
     \item We customize the DNN architecture to effectively exploit inherent patterns of wideband THz massive MIMO channels in the angle-distance-frequency domain. Furthermore, the attention mechanism is applied to make the DNN adaptive to different system configurations by dynamically re-weighting network features.
     \item We creatively design a weighted normalized mean-squared error (NMSE) loss function to realize effective mixed training of data from different system configurations, so that we can obtain a single robust DNN that works well under various configurations.
\end{itemize}

The rest of this paper is organized as follows. Section II introduces the system and channel model, as well as the problem formulation. In Section III, the proposed deep unfolding-based channel estimator is elaborated in detail, whose superiority is then demonstrated by numerical results provided in Section VI. Eventually, the paper is concluded in Section VII along with future work.

\emph{Notations:} We use italic, bold-face lower-case, and bold-face upper-case letters to denote scalar, vector, and matrix, respectively. $(\cdot)^{T}$, $(\cdot)^{H}$, $|\cdot|$, and $||\cdot||_F$ denote the transpose, conjugate transpose, modulus, and Frobenius norm, respectively. $\mathrm{diag(\cdot)}$ converts a vector to a diagonal matrix. $(\cdot)^{.2}$ and $./$ denote element-wise squaring and division, respectively. $\bm{1}_{a}$ and $\bm{0}_{a}$ denote the $(a,1)$-dimensional all-one vector and all-zero vector, respectively. $a:b:c$ denotes the arithmetic sequence vector starting from $a$ and ending at $c$ with common difference $b$. ${\mathbb C^{x \times y}}$ denotes the ${x \times y}$ complex space. $\mathcal{CN}(\mu,\sigma^2)$ denotes a circularly symmetric complex Gaussian (CSCG) random variable with mean $\mu$ and variance $\sigma^2$ while $\mathcal{CN}(\bm{\mu},\bm{\Sigma})$ denotes a CSCG random vector with mean $\bm{\mu}$ and covariance $\bm{\Sigma}$. $\mathcal{U}[a,b]$ denotes the uniform distribution between $a$ and $b$.

\section{System model}
In this section, the massive MIMO system is first presented. Then, the wideband THz near-field channel model is introduced, after which the channel estimation process is formulated as a classic MMV-CS problem.

\subsection{Massive MIMO system}
\begin{figure*}[!htb]
	\centering
	\includegraphics[scale=0.5]{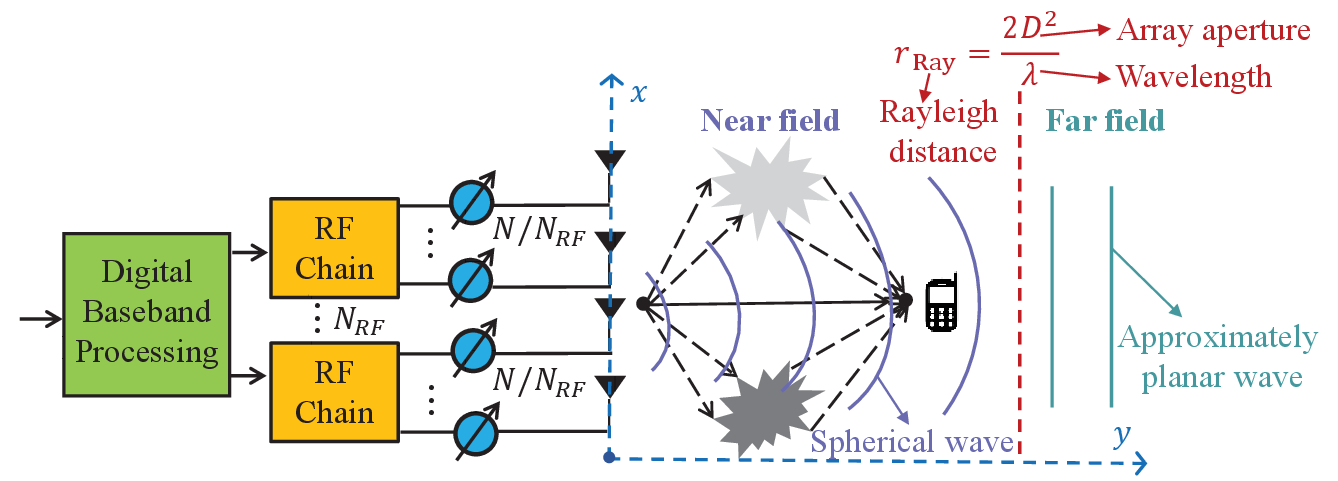}
	\caption{Partially-connected hybrid analog-digital massive MIMO system}
	\label{system}
\end{figure*}

As illustrated in Fig. \ref{system}, we consider that the base station (BS) equipped with an $N$-antenna uniform linear array (ULA) and $N_{RF}$ ($N_{RF}\ll N$) RF chains. To reduce the hardware overhead and power consumption, the partially-connected hybrid analog-digital architecture is considered, where each RF chain is connected to $N/N_{RF}$ antennas through $N/N_{RF}$ low-cost one-bit phase shifters. We consider the frequency division duplex (FDD) system where the BS transmits pilots to a single-antenna user\footnote{The extension to multiple multi-antenna users is straightforward since the pilots sent by the BS are broadcast to all users' antennas.} on $K$ subcarriers for downlink channel estimation. Once the channels are estimated at the user, they will be fed back to the BS for effective beamforming, which further facilitates accurate downlink signal decoding. Without loss of generality, we assume that the baseband pilot fed to all the RF chains is always $1$, then the received signal at the user on the $k$-th subcarrier in the $m$-th time slot can be expressed as
\begin{equation}
	y_{m}^k=\bm{w}_m^T\bm{h}^k+n_{m}^k,
\end{equation}
where $\bm{w}_m\in\mathbb{C}^{N\times 1}$, $\bm{h}^k\in \mathbb{C}^{N\times 1}$ and $n_{m}^k\sim \mathcal{CN}(0,\sigma^2)$ denote the phase shifts of the phase shifters at the BS, the channels from the BS to the user, and the additive noise with variance $\sigma^2$ at the user, respectively. Stacking the received signals of totally $M$ time slots within which the channel is assumed to be constant, we further obtain
\begin{equation}
	\bm{y}^k=\bm{W}\bm{h}^k+\bm{n}^k,
	\label{transmission_model}
\end{equation}
where $\bm{y}^k=[y_{1}^k,\cdots,y_{M}^k]^T\in \mathbb{C}^{M\times 1}$, $\bm{n}^k=[n_{1}^k,\cdots,n_{M}^k]^T\in \mathbb{C}^{M\times 1}$, and $\bm{W}=[\bm{w}_1,\cdots,\bm{w}_{M}]^T\in \mathbb{C}^{M\times N}$ whose elements are randomly selected from $\frac{1}{\sqrt{N}}\{+1,-1\}$ with equal probability.

\subsection{Wideband THz near-field channel model}
Compared to the low-frequency counterpart, the wideband massive MIMO channel model in the THz band is much more complicated due to the near-field beam split effect, which makes the array response vector dependent on angles, distances, and the frequency of the subcarrier. Denote $r$ as the distance from the center of the array at the BS to the scatter or the user and assume $r<r_{\text{Ray}}$, where $r_{\text{Ray}}=\frac{2D^2}{\lambda}$ denotes the Rayleigh distance \citep{near_field_dict} within which the near-field effect cannot be ignored. Furthermore, $D=(N-1)d$ denotes the array aperture with $d=\frac{\lambda}{2}$ denoting the antenna spacing, where $\lambda=\frac{c}{f_c}$ denotes the carrier wavelength with $c$ and $f_c$ denoting the speed of light and the central frequency, respectively.

Straightforwardly, the near-field array response vector at frequency $f$ can be expressed as
\begin{equation}\label{array_1}
\begin{aligned}
    & \bm{a}_N(f,r,r^{(0)},\cdots,r^{(N-1)})=\\
    & \frac{1}{\sqrt{N}}\left[e^{-j2\pi \frac{c}{f}(r^{(0)}-r)},\cdots,e^{-j2\pi \frac{c}{f}(r^{(N-1)}-r)}\right]^T,
\end{aligned}
\end{equation}
where the distance from the $n$-th antenna of the array to the scatter or the user, $r^{(n)}$, can be calculated according to the geometry as $r^{(n)}=\sqrt{r^2-2r\delta_nd\theta+\delta_n^2d}$ with $\theta\in[-1,1]$ denoting the sine of the angle of departure (AoD) at the BS and $\delta_n=\frac{2n-N+1}{2},n=0,1,\cdots,N-1$. Based on the Fresnel approximation, we further have $r^{(n)}\approx r-\delta_nd\theta+\frac{\delta_n^2d^2(1-\theta^2)}{2r}$ \citep{near_field_dict}. Therefore, (\ref{array_1}) can be rewritten as
\begin{equation}\label{array_2}
\begin{aligned}
    & \bm{a}_N(\theta,r,f)=\frac{1}{\sqrt{N}}\Big[e^{-j2\pi\frac{c}{f}\left(\frac{\delta_0^2d^2(1-\theta^2)}{2r}-\delta_0d\theta\right)},\\
    & \cdots,e^{-j2\pi\frac{c}{f}\left(\frac{\delta_{N-1}^2d^2(1-\theta^2)}{2r}-\delta_{N-1}d\theta\right)}\Big]^T.
\end{aligned}
\end{equation}

Using (\ref{array_2}), the $k$-th downlink subchannel can be expressed as
\begin{equation}
	\bm{h}^k=\sqrt{\frac{N}{N_cN_p}}\sum_{i=1}^{N_c}\sum_{j=1}^{N_p}\alpha_{i,j}e^{-j2\pi f_k \tau_{i,j}}\bm{a}_N(\theta_{i,j},r_{i,j},f_k),
	\label{channel_model}
\end{equation}
where $N_c$ and $N_p$ denote the number of clusters and the number of subpaths in a cluster, respectively. Besides, $\alpha_{i,j},\tau_{i,j},\theta_{i,j}$ denote the path gain, the delay, and the sine of the AoD of the $i$-th subpath in the $j$-th cluster, respectively, while $r_{i,j}$ denotes the distance from the center of the BS's array to the scatter or the user corresponding to the $i$-th subpath in the $j$-th cluster. Furthermore, we have $f_k=f_c+(k-1-\frac{K-1}{2})\frac{f_s}{K}$ with $f_k$ and $f_s$ denoting the frequency of the $k$-th subcarrier and the bandwidth, respectively, and $\theta_{i,j}=\text{sin}(\phi_{i,j})$ with $\phi_{i,j}$ denoting the physical AoD of the $i$-th subpath in the $j$-th cluster.

\subsection{Formulation of the MMV-CS problem}
\begin{figure}[!htb]\small
\centering
\begin{tabular}{c}
\includegraphics[scale=0.5]{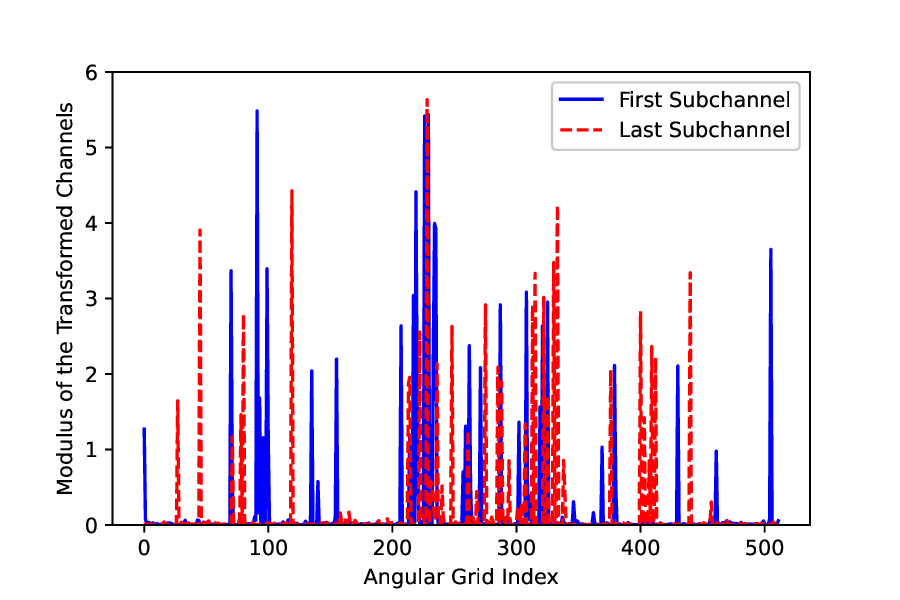}\\
{\footnotesize\sf (a) Use a common angular dictionary.} \\
\includegraphics[scale=0.5]{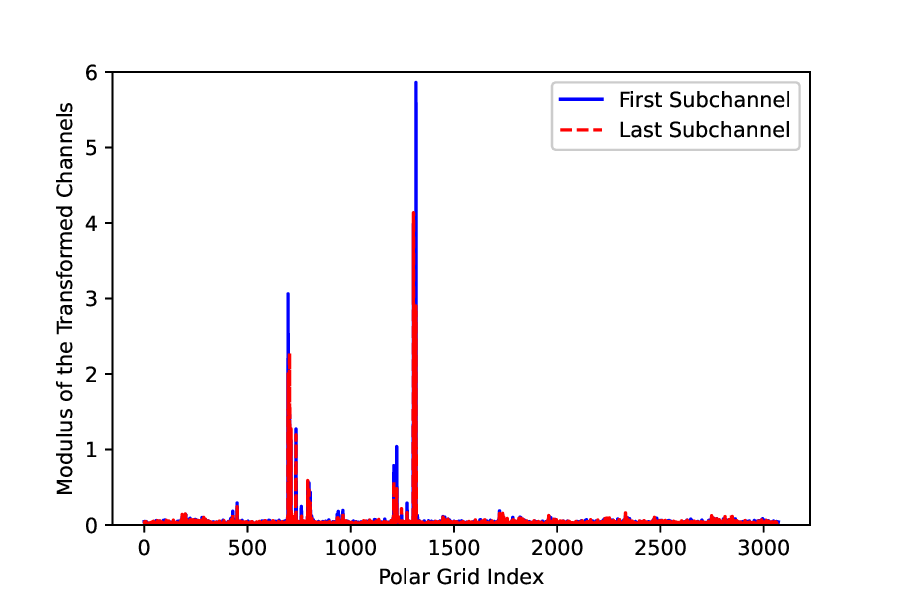}\\
{\footnotesize\sf (b) Use frequency-dependent polar dictionaries.} \\
\end{tabular}
\caption{Impact of dictionaries under the default simulation setting}
\label{impact_of_dict}
\end{figure}

To facilitate channel estimation, it is a common practice to first perform domain transformation with a proper dictionary in massive MIMO systems so that channels in the transformed domain possess appealing properties such as sparsity and sparsity structures to help solve the under-determined equation in (\ref{transmission_model}). To deal with the aforementioned near-field beam split effect in wideband THz channels, the dictionary should be designed as follows. On one hand, the atoms of the dictionary should naturally conform to the form of the near-field array response vector to enhance channel sparsity in the transformed domain \citep{near_field_dict}. On the other hand, instead of using a common dictionary on all subcarriers, frequency-dependent dictionaries should be applied on different subcarriers to compensate for the beam split effect similar to the far-field case \citep{uamp_sbl_unfolding}. Specifically, the dictionary for the $k$-th subchannel is designed as
\begin{equation}
	\bm{A}^k=[\bm{A}^k_0,\cdots,\bm{A}^k_{S-1}]\in \mathbb{C}^{N\times G},
	\label{dicts}
\end{equation}
where $S$ denotes the number of sampled distance grids and the $s$-th sub-dictionary is further composed of $Q$ near-field array response vectors as
\begin{equation}
	\bm{A}^k_s=[\bm{a}_N(\theta_0,r_{s,0},f_k),\cdots,\bm{a}_N(\theta_{Q-1},r_{s,Q-1},f_k)],
\end{equation}
where $Q$ denotes the numbers of sampled angle grids. So, the total number of grids in the transformed polar domain is $G=SQ$. According to \citep{near_field_dict}, to minimize the maximum coherence between arbitrary two atoms to improve the performance of CS algorithms, the angles should be uniformly sampled while the distances should be non-uniformly sampled as
\begin{equation}
	\theta_q=\frac{2q-Q+1}{Q},q=0,\cdots,Q-1,
\end{equation}
\begin{equation}
	r_{s,q}=\frac{1}{s}Z_{\triangle}(1-\theta_q^2),s=0,\cdots,S-1,
\end{equation}
where $Z_{\triangle}=\frac{N^2d^2}{2\beta_{\triangle}^2c/f_c}$ with $\beta_{\triangle}$ denoting the threshold that balances coherence level and grid resolution, and $S$ is set to the minimum integer that satisfies $\frac{1}{S}Z_{\triangle}<\rho_{\text{min}}$ with $\rho_{\text{min}}$ denoting the minimum allowable distance \citep{near_field_dict}. Using the dictionary in (\ref{dicts}), we can readily perform domain transformation as
\begin{equation}
	\bm{h}_{k}\approx\bm{A}^k\bm{x}^k,
	\label{domain_transform}
\end{equation}
where $\bm{x}^k\in \mathbb{C}^{G\times 1}$ denotes the $k$-th sparse polar domain subchannel and the $\approx$ is due to the quantization error caused by the finite grid resolution, which is usually very small in massive MIMO systems with dense grids \citep{sbl_unfolding}. As shown in Fig. \ref{impact_of_dict}, compared to using a common angular dictionary, using frequency-dependent polar dictionaries not only enhances the channel sparsity but also recovers the common sparsity structure among subchannels, thus effectively compensating for the near-field beam split effect.

Taking the $\approx$ in (\ref{domain_transform}) by $=$ and substitute it into (\ref{transmission_model}), we can obtain the following CS model \begin{equation}
	\bm{y}^k=\bm{\Phi}^k\bm{x}^k+\bm{n}^k,\forall k,
	\label{signal_model_vector}
\end{equation}
where $\bm{\Phi}^k\triangleq\bm{WA}^k$ denotes the measurement matrix on the $k$-th subcarrier. In a word, the goal of channel estimation is to accurately recover $\bm{x}^k,\forall k$ based on $\bm{y}^k,\bm{\Phi}^k,\forall k$, and $\sigma^2$. Since we have $K$ measurements on $K$ subcarriers and $\bm{x}^k,\forall k$ have basically the same sparse supports, i.e., the common sparsity structure, it is a typical MMV-CS problem. After the sparse polar domain subchannels are estimated, the original subchannels can be readily reconstructed by (\ref{domain_transform}).

\section{Deep unfolding-based channel estimator}
In this section, the principles of the AMP-SBL algorithms are first introduced briefly. Then, the proposed deep unfolding-based algorithm that enhances the capability of AMP-SBL by DL is elaborated in detail, including its architecture, training scheme, and gains.
\subsection{AMP-SBL}
As one of the most powerful CS algorithms, SBL has high theoretical sparse recovery performance and is flexible to exploit various sparsity structures \citep{sbl_ce}. Omitting the subcarrier superscript, to recover a particular sparse polar domain subchannel $\bm{x}$, SBL first assumes that
\begin{equation}
	\bm{x}\sim \mathcal{CN}(\bm{0}_G,\mathrm{diag}(\bm{\gamma})),
\end{equation}
where $\bm{\gamma}=[\gamma_1,\cdots,\gamma_{G}]^T$ and $\gamma_g$ denotes the Gaussian variance parameter of the $g$-th element of $\bm{x}$. Then, dozens of expectation-maximization (EM)-based iterations are executed. In each iteration, the E-step computes the posterior mean and covariance of $\bm{x}$ given $\bm{y}$, based on which $\bm{\gamma}$ is updated by the M-step. After convergence, the eventual posterior mean of $\bm{x}$ is regarded as an estimate of it. The proven sparse-promoting property of SBL ensures that $\bm{\gamma}$ will gradually become a sparse vector as the iteration progresses \citep{sbl_ce}, i.e., most elements in $\bm{\gamma}$ will be close to zero after convergence and the indices of those non-zero elements indicate the existence of channel paths at the corresponding angle and distance.

Since the E-step in SBL involves the computation of matrix multiplications and inversions, its complexity is very high, especially in the THz massive MIMO channel estimation problem where the dimensions of matrices are quite large. To alleviate this issue, the AMP-SBL algorithm in \citep{uamp_sbl} uses an alternative realization of the E-step based on AMP. Since only matrix-vector multiplications are involved, the complexity is dramatically reduced compared to the original realization of the E-step. Furthermore, unitary preprocessing is applied to improve the robustness to general measurement matrices. The extension of AMP-SBL to the MMV case has been also straightforwardly made in \citep{uamp_sbl}, where a common Gaussian variance vector is adopted and updated according to information from all measurements to exploit the common sparsity structure. The detailed algorithm is demonstrated in {\bfseries Algorithm 1}, where $L$ denotes the number of iterations, lines $1-2$ represent the unitary preprocessing where singular value decomposition is executed on each measurement matrix, line $3$ represents the initialization procedure, lines $5-7$ represent the AMP-based E-step executed on all subcarriers in the $l$-th iteration, and line $8$ represents the M-step in the $l$-th iteration. Notice that, we adopt frequency-dependent measurement matrices here, causing some slight differences from the original algorithm proposed in \citep{uamp_sbl}.
\begin{algorithm*}
	\caption{AMP-SBL for MMV-CS}
	\textbf{Input}: $\bm{y}^k,\bm{\Phi}^k,\forall k;\sigma^2,L$
	\begin{algorithmic}[1]
		\STATE $\forall k: \bm{\Phi}^k=\bm{U}^k\bm{\Sigma}^k\bm{V}^k$ 
		\STATE $\forall k: \bm{r}^k=(\bm{U}^k)^H\bm{y}^k$, $\bm{A}^k=(\bm{U}^k)^H\bm{\Phi}^k$ 
		\STATE $\bm{\mu}_{\bm{x}^k}^0=\bm{0}_{G},(\bm{s}^k)^{0}=\bm{0}_{Q},\forall k;\bm{\gamma}^0=\bm{1}_{G},\epsilon^0=0.001$ 
		\STATE \textbf{for} $l=1:1:L$ \textbf{do}
		\STATE \quad $\forall k: \bm{\tau}_p^k = |\bm{A}^k|^{.2}\bm{\tau}_{\bm{x}^k}^{l-1},\ \bm{p}^k=\bm{A}^k\bm{\mu}_{\bm{x}^k}^{l-1}-\bm{\tau}_p^k\cdot (\bm{s}^k)^{l-1},\ \bm{\tau}_s^k=\bm{1}./(\bm{\tau}_p^k+\sigma^2\bm{1})$
		\STATE \quad $\forall k: (\bm{s}^k)^l=\bm{\tau}_s^k\cdot(\bm{r}^k-\bm{p}^k),\ \bm{\tau}_q^k = \bm{1}./\left(\left|\left(\bm{A}^k\right)^H\right|^{.2} \bm{\tau}_s^k\right), \ \bm{q}^k=\bm{\mu}_{\bm{x}^k}^{l-1}+\bm{\tau}_q^k\cdot\left(\left(\bm{A}^k\right)^H\left(\bm{s}^k\right)^l\right)$
		\STATE \quad $\forall k: \bm{\mu}_{\bm{x}^k}^l=\bm{q}^k./(\bm{1}+\bm{\tau}_q^k\cdot \bm{\gamma}^{l-1}),\ \bm{\tau}_{\bm{x}^k}^l=\bm{\tau}_q^k./(\bm{1}+\bm{\tau}_q^k\cdot \bm{\gamma}^{l-1})$
		\STATE \quad $\bm{\gamma}^l=\frac{2\epsilon^{l-1}+1}{\frac{1}{K}\sum_{k=1}^K(|\bm{\mu}_{\bm{x}^k}^l|^2+\bm{\tau}_{\bm{x}^k}^l)},\ \epsilon^l=\frac{1}{2}\sqrt{\text{log}_{10}\left(\frac{1}{G}\sum_{g=1}^G\gamma_g^l\right)-\frac{1}{G}\sum_{g=1}^G\text{log}_{10}\gamma_g^l}$ 
		\STATE \textbf{end for}
	\end{algorithmic}
	\textbf{Output}: $\hat{\bm{x}}^k=\bm{\mu}_{\bm{x}^k}^L,\forall k$
\end{algorithm*}

\subsection{AMP-SBL unfolding}
In practice, although the unitary preprocessing improves the robustness to some extent, the AMP-SBL algorithm can still easily diverge under structured measurement matrices, such as those adopted in this paper, making its low complexity meaningless. On the other hand, the M-step in the SBL algorithm is derived based on the assumption that the elements of $\bm{x}$ are independent of each other \citep{sbl_ce}, which is not true in practical THz channels. With clusters and power leakage among grids \citep{sbl_unfolding}, the elements of $\bm{x}$ corresponding to close polar grids tend to have close modulus, which results in the block sparsity structure. In this case, the original M-step is far from optimal.

To overcome the shortcomings of AMP-SBL and obtain a channel estimation algorithm with good performance, low complexity, and strong robustness, we propose a deep unfolding-based algorithm. Specifically, after the unitary preprocessing procedure, the EM-based iterations are unfolded into a cascaded large DNN, whose each layer, named ``AMP-SBL layer", corresponds to an iteration of the AMP-SBL algorithm. In each AMP-SBL layer, the AMP-based E-step remains unchanged while the original simple M-step is replaced by a complicated function realized by a small DNN, whose architecture is carefully designed to effectively exploit the inherent channel patterns.
\begin{figure*}[htb!]
	\centering
	\includegraphics[scale=0.5]{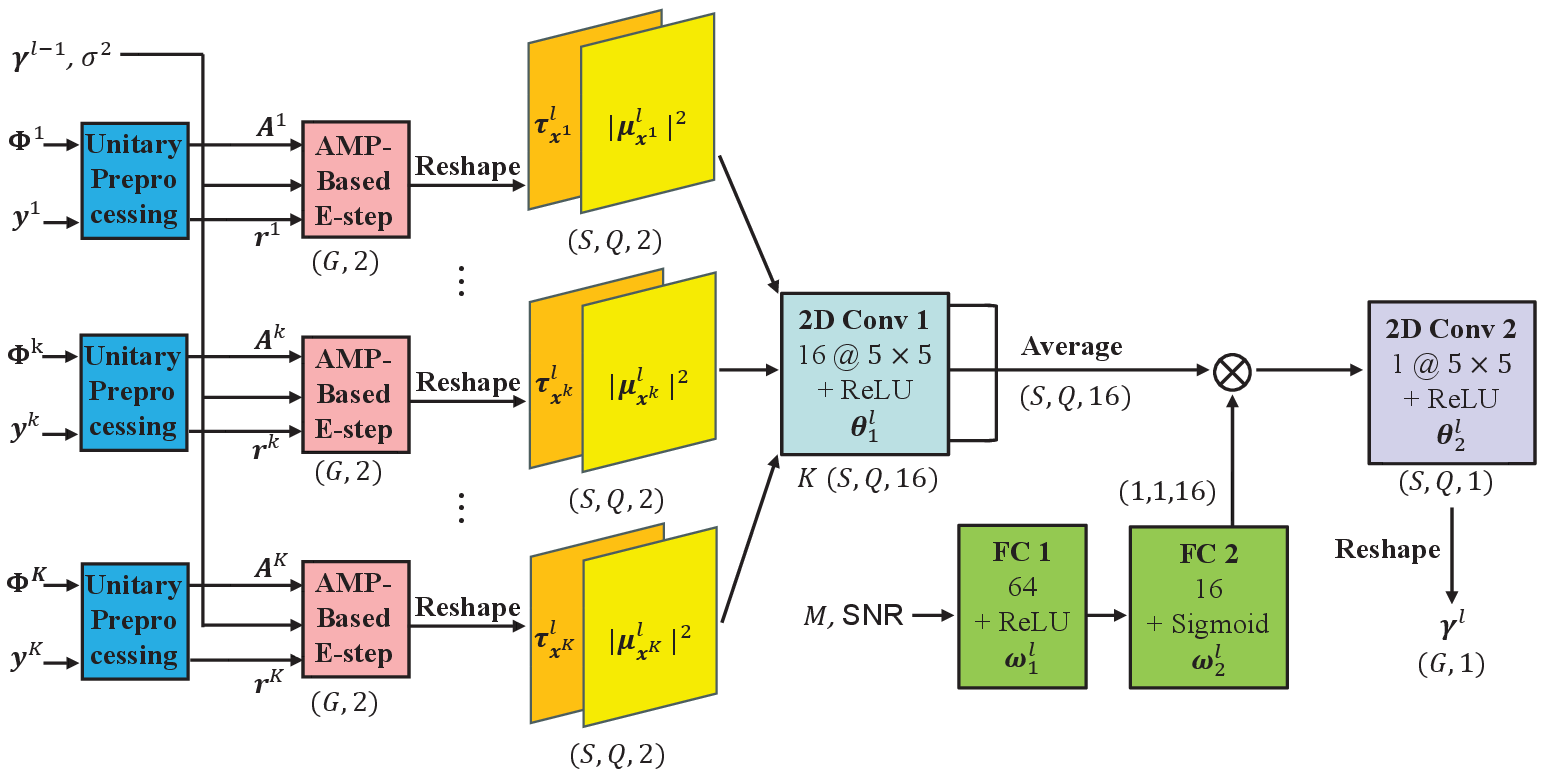}
	\caption{The architecture of the $l$-th AMP-SBL layer}
	\label{network}
\end{figure*}

As illustrated in Fig. \ref{network}, in the $l$-th AMP-SBL layer, the AMP-based E-step is first executed on all $K$ subcarriers, where $|\bm{\mu}_{\bm{x}^k}^l|^2,\bm{\tau}_{\bm{x}^k}^l$ are obtained based on the common $\bm{\gamma}^{l-1},\sigma^2$ and the subcarrier-specific $\bm{A}^k,\bm{r}^k$ on the $k$-th subcarrier. After that, for each $k$, $|\bm{\mu}_{\bm{x}^k}^l|^2,\bm{\tau}_{\bm{x}^k}^l$ are reshaped into $(S,Q)$-dimensional matrices since the block sparsity structure exists in the 2D polar domain and are further stacked along the last expanded dimension to obtain the $(S,Q,2)$-dimensional feature tensor. Then, the feature tensor is processed by a 2D convolutional (Conv) layer with $16$ filters of size $5\times 5$. Zero padding is executed to keep the dimensions unchanged after convolution. To exploit the common sparsity structure, $K$ different $(S,Q,16)$-dimensional tensors are then averaged to obtain a single $(S,Q,16)$-dimensional tensor. Eventually, another 2D Conv layer with a single filter of size $5\times 5$ is used to output the $(S,Q,1)$-dimensional updated variance parameter tensor, which is reshaped back to a $(G,1)$-dimensional vector $\bm{\gamma}^l$ to facilitate the next AMP-SBL layer's computation.

Most existing DNN-based channel estimators need to be trained separately under different system configurations and the lack of robustness dramatically reduces their practical values. In contrast, the proposed algorithm naturally works with different system scales such as the number of time slots $M$ since the dimensions of the feature maps going through the Conv layers, i.e., $S$ and $G$, are manually selected hyperparameters, thus can be fixed. Furthermore, we add an attention module to the backbone network to enhance its adaptability to different configurations. Specifically, two fully-connected (FC) layers are used to predict $16$ weights according to configuration parameters, which are then multiplied to the feature maps before the second Conv layer in the backbone network along the last dimension. Notice that, although we only consider $M$ and SNR here as an example, many other configurations parameters can also be included in practice. Through such dynamic feature re-weighting process, the DNN can adapt to different configurations flexibly by changing the attentions paid to different features \citep{attention_CE}. Overall, the function of the DNN-based M-step in the $l$-th AMP-SBL layer can be mathematically expressed as (\ref{update_gamma_uamp_residual}), where $f^l_{C_1}(\cdot;\bm{\theta}^l_1)$ and $f^l_{C_2}(\cdot;\bm{\theta}^l_2)$ denote the operations of the first and the second Conv layers with weights $\bm{\theta}^l_1$ and $\bm{\theta}^l_2$, respectively, and $f^l_{D_1}(\cdot;\bm{w}^l_1)$ and $f^l_{D_2}(\cdot;\bm{w}^l_2)$ denote the operations of the first and the second FC layers with weights $\bm{w}^l_1$ and $\bm{w}^l_2$, respectively. $g(x)=\mathrm{max}(0,x)$ denotes the ReLU activation function to enhance the DNN's representation ability or guarantee the non-negativity of the updated Gaussian variances, and $\delta(x)=1/(1+e^x)$ denotes the Sigmoid activation function to generate attention weights between $0$ and $1$.
\begin{figure*}[b]
    \begin{align}\label{update_gamma_uamp_residual}
\bm{\gamma}^l=g\left(f^l_{C_2}\left(\delta\left(f_{D_2}^l\left(g\left(f_{D_1}^l\left(M,\text{SNR};\bm{w}_1^l\right)\right);\bm{w}_2^l\right)\right)\cdot\frac{1}{K}\sum_{k=1}^Kg\left(f^l_{C_1}\left(|\bm{\mu}_{\bm{x}^k}^l|^2,\bm{\tau}_{\bm{x}^k}^l;\bm{\theta}^l_1\right)\right);\bm{\theta}^l_2\right)\right)
    \end{align}
\end{figure*}

\subsection{Training scheme}
The optimal $\bm{\theta}^l_1,\bm{\theta}^l_2,\bm{w}^l_1,\bm{w}^l_2,\forall l$ need to be obtained through training. We generate $8,000$, $1,000$, $1,000$ channel samples with various system configurations as the training set, the validation set, and the testing set, respectively. Due to the stacking of multiple network layers and activation functions, the loss function is nonconvex and can not be guaranteed to converge to the global optimum. To achieve good performance, the following training techniques are adopted. First of all, the Adam optimizer is adopted to exploit both first- and second-order momentums. Besides, to avoid getting stuck in bad local optimum, layer-wise training is adopted \citep{uamp_sbl_unfolding}, where shallower networks are trained first and deeper networks with newly added layers are trained on their basis. Last but not least, the weights of each newly added layer are initialized by its previous layer's weights, which reduces the loss oscillation and accelerates the convergence speed dramatically. When the performance does not improve with a new layer added, the previous network without the newly added layer is the final algorithm. In simulation, the typical value of $L$ is $10$. Besides, the initial learning rate is set to $10^{-3}$, and strategies including learning rate decay and early stopping are used in each training to improve training speed and prevent overfitting. The batch size is set to $16$ due to the memory limit of the used GPU.

Since the channel estimation performance under different system configurations, such as $M$ and SNR, varies a lot, configurations with better performance will be overwhelmed by configurations with worse performance during mixed training if we simply adopt NMSE as the loss function, resulting in unbalanced training levels of different configurations \citep{attention_CE}. To deal with this issue, we propose the following weighted NMSE loss function:
\begin{equation}
	\label{loss}
	\text{Loss}(\bm{H},\hat{\bm{H}},M,\text{SNR})=a(M,\text{SNR})\frac{||\bm{H}-\hat{\bm{H}}||_F^2}{||\bm{H}||_F^2},
\end{equation}
where $\bm{H}\triangleq[\bm{h}_1,\cdots,\bm{h}_K]$ denotes true subchannels while $\hat{\bm{H}}$ denotes the subchannels output by the proposed algorithm. The weight corresponding to configuration parameters $M$ and SNR, $a(M,\text{SNR})$, is set to the ratio of the SBL algorithm's NMSE under a middle configuration point and that under $M,\text{SNR}$, so that smaller weights are assigned to the losses of harsher configurations while larger weights are assigned to the losses of easier configurations. In this way, the loss levels of different configurations are balanced so that all configurations can get sufficient training, leading to effective mixed training.

\subsection{Gain analysis}
Compared to existing CS-based and DL-based algorithms, the proposed algorithm has gains in the following aspects, which will also be validated by simulation results later.
\begin{itemize}
\item First, since the approximation loss of the AMP-based E-step can be effectively compensated by the properly trained DNN, the proposed algorithm can converge under structured measurement matrices where the original AMP-SBL algorithm diverges, thus making its low complexity meaningful.
\item Apart from that, the proposed algorithm is also promising to achieve better performance than SBL since the DNN can exploit the actual channel distribution and learn a matched function of the M-step, which includes the original M-step as a special case. Besides, the solid foundation of the advanced SBL algorithm makes the proposed approach more potential for great performance than other DL-based methods that are based on simple basic algorithms or purely data-driven.
\item In terms of complexity, AMP simplifies the E-step per iteration while the DNN-based M-step dramatically improves the convergence speed at the cost of moderate extra complexity. Therefore, the proposed approach has lower complexity than SBL and AMP-SBL, and its complexity is close to the simple SOMP \citep{near_field_dict}. Specifically, the numbers of real floating operations (FLOPs) of SBL, AMP-SBL, the proposed algorithm, and SOMP are $16KM^2L_1G$, $20KML_2G$, $(20KM+800)LG$, and $8KML_3G$ respectively, where $L_1$, $L_2$, $L_3$ denote the numbers of iterations of SBL, AMP-SBL, and SOMP, respectively.
\item Last but not least, the numbers of required training samples and model parameters are quite small thanks to the model-driven nature. Besides, the proposed mixed training method can dramatically reduce the DNN training and storage overhead at the user side in FDD systems without sacrificing performance compared to separate training. Specifically, a $N_{\text{config}}$-fold reduction where $N_{\text{config}}$ denotes the number of possible configurations. Such a robust and consistent model is also simple to deploy and can reduce latency in fast-changing scenarios.
\end{itemize}

\section{Simulation results}
\label{simulation}
In this section, simulation results\footnote{For reproduction, source code is available at $\mathrm{\textcolor{blue}{https://github.com/EricGJB/Deep\_Unfolding\_THz\_CE}}$} are provided to validate the superiority of the proposed algorithm. The default setting is as follows unless specified: $N=256$, $K=32$, $f_c=100$ GHz, and $f_s=10$ GHz, so that $r_{\text{Ray}}=97.5375\text{m}$. Besides, $Q=512$, $\beta=1.2$, $\rho_{\text{min}}=3\text{m}$ \citep{near_field_dict}, so that $S=6$ and $G=3072$. $N_c=3$, $N_p=10$, $\alpha_{i,j}\sim \mathcal{CN}(0,1)$, the angles and distances of subpaths within a cluster are assumed to obey the Laplacian distributions whose means obey $U[0^\circ,360^\circ]$ and $U[5\text{m},30\text{m}]$, respectively while the standard deviations are set to $4^\circ$ and $1\text{m}$, respectively \citep{NB_OMP,THz_channel_model}. The signal-to-noise-ratio (SNR) is defined as $1/\sigma^2$ and the middle configuration point is $M=48,\text{SNR}=10$ dB. All the performance points are obtained by averaging over 200 random data samples.

For benchmarks, we mainly compare with two on-grid CS algorithms, namely the SOMP algorithm \citep{near_field_dict,NB_OMP} and the MMV version of the original SBL algorithm, MSBL \citep{sbl_ce,sbl_unfolding}, as well as their counterparts with different dictionaries and training schemes. To focus on highlighting the unique contributions of this paper, off-grid CS algorithms and other DL-based algorithms are not compared, since the former usually have poor performance in cluster channels while the latter has shown obvious inferiority to the proposed approach in the mmWave band\citep{sbl_unfolding,uamp_sbl_unfolding}.

\begin{figure}[!htb]
	\centering
	\includegraphics[scale=0.5]{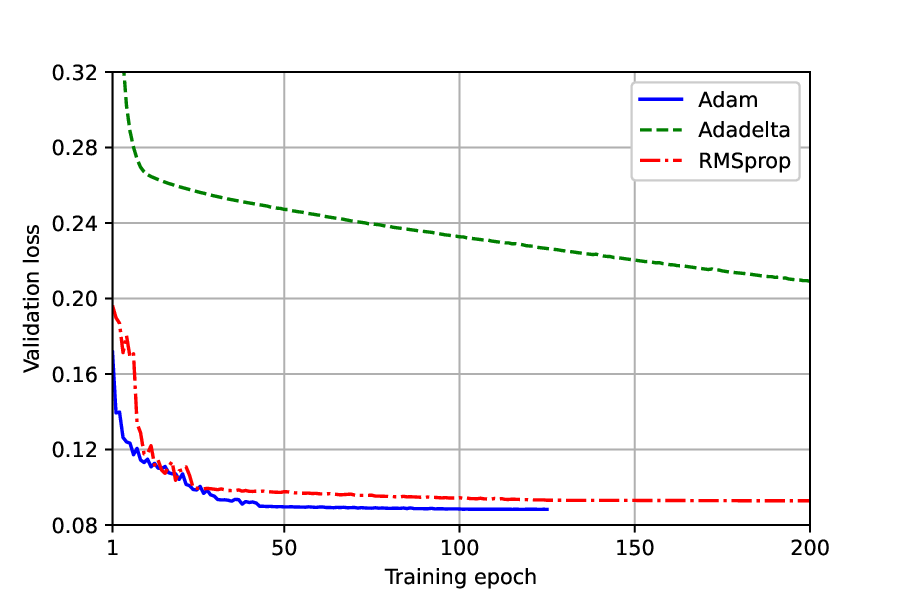}
	\caption{Impact of optimizer on DNN training. Three AMP-SBL layers are unfolded.}
    \label{impact_of_optimizer}
\end{figure}

First of all, we would like to clarify that the hyperparameters in the proposed approach, including the network architecture, the optimizer, the learning rate, and so on, are determined through experiments. Specifically, different hyperparameters are tried and the combination with the lowest validation loss is selected. For instance, Fig. \ref{impact_of_optimizer} shows the impact of optimizer on DNN training. As we can see, Adam outperforms other optimizers in terms of convergence speed and performance. Due to limited space, the determination process of other hyperparameters are omitted here.

\begin{figure}[htb!]
	\centering
	\includegraphics[scale=0.5]{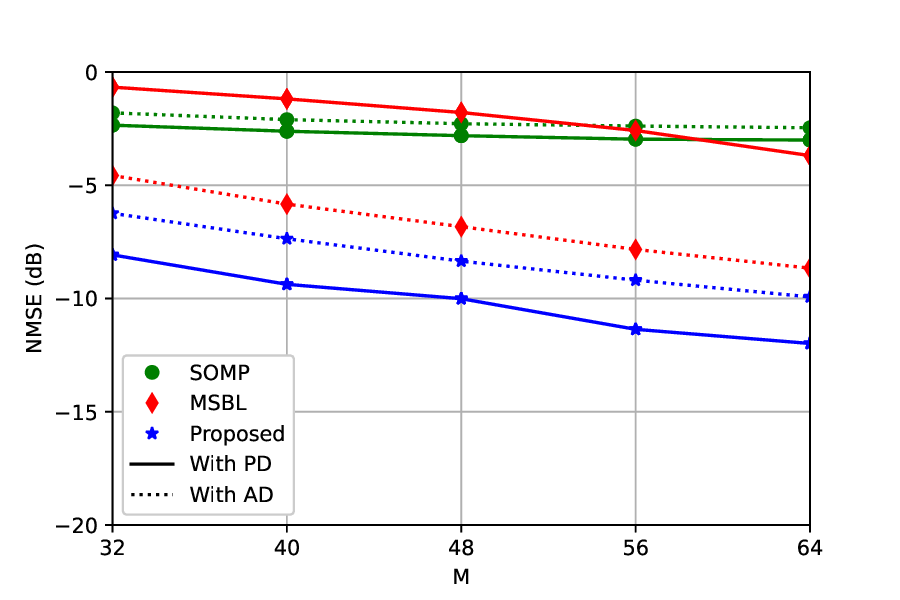}
	\caption{NMSE versus $M$ with different dictionaries}
	\label{impact_of_M}
\end{figure}

Fig. \ref{impact_of_M} shows the impacts of dictionary under different $M$, where in different dictionaries are distinguished by line types. First of all, larger $M$ reasonably leads to better estimation performance of all algorithms thanks to more information gathered about the channels while the pilot overhead grows as well. Although the polar dictionaries (PD) leads to better performance than the angular dictionaries (AD) in SOMP thanks to stronger sparsity, the situation is reversed in MSBL due to larger coherence among atoms. However, in the proposed deep unfolding algorithm, the performance superiority of the PD appears again thanks to the compensation effect of the DNN, resulting in the lowest NMSE among all algorithms under various $M$. This kind of performance superiority can also help save pilot overhead. For instance, to achieve a $-9$ dB NMSE, MSBL-AD requires $M=64$ while AMP-SBL unfolding-PD only requires $M=37$. At last, notice that AMP-SBL is not visible in the figure since it has a high probability of divergence and thus terrible average NMSE under structured measurement matrices, which reflects the vital role of DNN in the proposed approach.

\begin{figure}[htb!]
	\centering
	\includegraphics[scale=0.5]{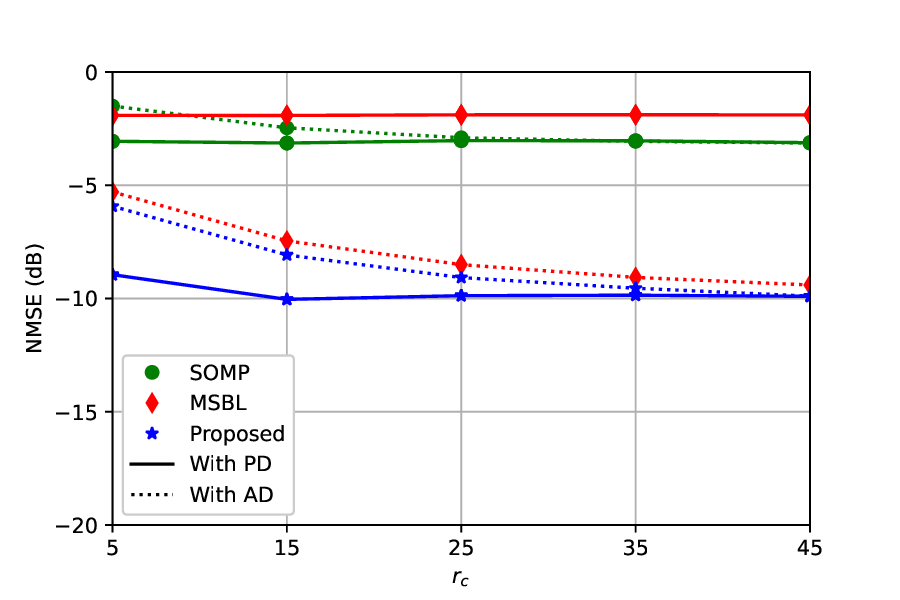}
	\caption{NMSE versus $r_{c}$}
	\label{impact_of_r}
\end{figure}

Furthermore, the behaviors of different algorithms under different levels of near-field beam split are investigated. Fig. \ref{impact_of_r} illustrates the impact of distance, wherein all subpaths' distances are set to a common value for convenient comparison, i.e., $r_{i,j}=r_{c},\forall i,j$. Again, PD leads to better performance in SOMP and the proposed algorithm while AD is better in MSBL. When $r_c$ is small, the performance gaps between using PD and AD are large in both SOMP and the proposed algorithm due to the relatively strong near-field effect. As $r_c$ increases, the performance gaps gradually narrow and eventually vanish when $r_c$ is large enough. As for the impact of bandwidth, since the beam split effect is compensated by the frequency-dependent measurement matrices and the common sparsity structure among subchannels always holds, the channel estimation performance of various algorithms rarely changes with the bandwidth.

Apart from better performance, the proposed approach also has much lower complexity than MSBL with only about $\frac{1}{37}$ per-iteration FLOPs and $\frac{1}{10}$ iteration number, as shown in Table \ref{complexity}. Thanks to its DNN-based implementation, the proposed approach even runs faster than SOMP on the same CPU. Besides, with more iterations unfolded and more layers, more and larger kernels used in the DNN in each iteration, the performance of the proposed approach will gradually improves with stronger representation ability. However, the complexity increases at the same time. Finally, the performance curve will reach a plateau where further increasing the network scale won't lead to performance improvement and may even risk overfitting. In this paper, hyperparameters at the turning point of the performance curve are selected to achieve the best performance with the least complexity. Nevertheless, it is totally feasible to adopt smaller network scales in practice to sacrifice performance to meet strict complexity requirements.

\begin{table}[h]	
	\centering
	\begin{tabularx}{0.5\textwidth}{|c|X|X|X|X|}
		\hline
		Algorithm & FLOPs per iteration ($\times10^9$) & Number of iterations & Overall Flops ($\times10^9$) & Average running time (ms)\\ \hline		
        SOMP-PD & 0.037 & 6 & 0.226 & 1.23 \\ \hline
        MSBL-AD & 3.623 & 100 & 362.388 & 415.69 \\ \hline
        Proposed & 0.097 & 10 & 0.978 & 0.89 \\ \hline
	\end{tabularx}
	\caption{Complexity comparison among different algorithms.}
	\label{complexity}
\end{table}

\begin{figure}[htb!]
	\centering
	\includegraphics[scale=0.5]{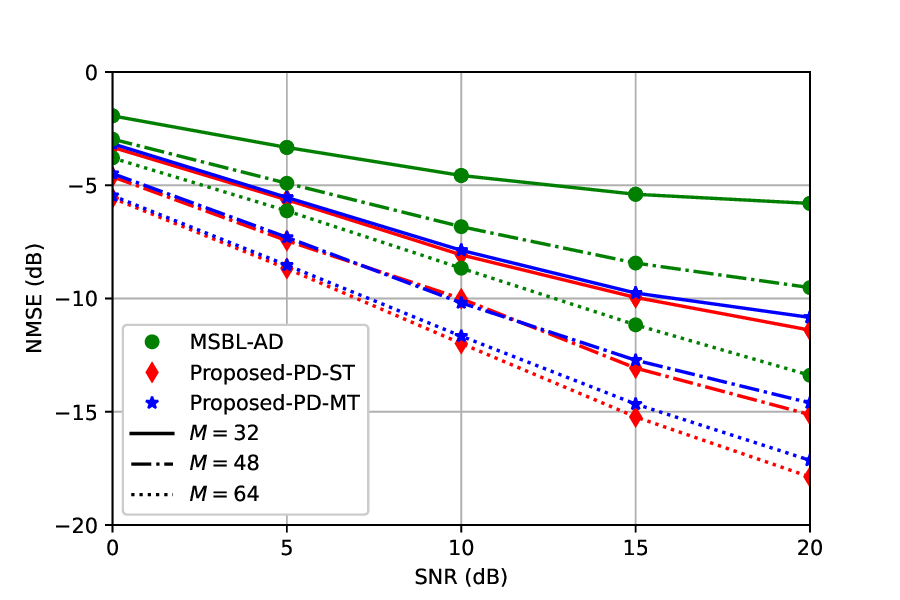}
	\caption{NMSE under different system configurations}
	\label{impact_of_snr}
\end{figure}

Eventually, to verify the generality and robustness of the proposed algorithm, Fig. \ref{impact_of_snr} shows the performance of different algorithms with different training schemes under various system configurations, including SNR and $M$ (distinguished by line types). As can be seen from the figure, the performance superiority of the proposed algorithm over MSBL stands under various system configurations, validating its generality. Compared to training a bunch of DNNs separately (ST) in different configurations, the proposed attention mechanism and the weighted NMSE loss function can jointly realize effective mixed training (MT) of data from different configurations with only very slight performance degradation at some configuration points, which is reflected by the close curves of ST and MT in the figure. Specifically, the original unnormalized NMSEs of the proposed approach at different configuration points differ by at most $28.5399$ and will lead to unbalanced mixed training. Applying the weights calculated based on MSBL-AD's NMSEs, the NMSE difference factor reduces to only $2.0392$, and the NMSEs of the proposed approch at all configuration points are basically at the same level around $0.1$. Besides, the data number of MT equals that of each configuration point in ST, so the total data amount of MT is much smaller than that of ST.

\section{Conclusion and future work}
In this paper, we propose a deep unfolding-based algorithm for wideband THz near-field massive MIMO channel estimation. We first compensate for the near-field beam split effect by using frequency-dependent near-field domain transformation dictionaries. Then, we enhance the capability of the AMP-SBL algorithm by using a DNN to learn the optimal parameter update rule in each of its iteration. The DNN architecture is customized to exploit inherent channel patterns. Finally, we propose an effective mixed training method based on novel designs of the DNN architecture and the loss function to obtain a single robust network that can work under various configurations. Simulation results demonstrate the good performance, low complexity, and strong robustness of the proposed algorithm. In the future, we will apply the proposed algorithm to more wireless communications problems and prove the convergence property to improve its universality and reliability.

\section*{Contributors}
Jiabao GAO performed the simulations and drafted the paper. Xiaoming CHEN and Geoffrey Ye LI helped organize, revise, and finalize the paper.

\section*{Compliance with ethics guidelines}
Xiaoming CHEN is a corresponding expert of Frontiers of Information Technology \& Electronic Engineering. Jiabao GAO, Xiaoming CHEN, and Geoffrey Ye LI declare that they have no conflict of interest.


\bibliographystyle{fitee}
\bibliography{bibsample}

\end{document}